# La structure du capital et la profitabilité
# Le cas des entreprises industrielles françaises

Mazen KEBEWAR[*]

Résumé

L'objectif de cet article est d'analyser l'impact de la structure du capital sur la profitabilité. Cet impact peut être expliqué par trois théories essentielles: la théorie du signal, l'influence de la fiscalité et la théorie de l'agence. Nous montrons, à partir d'un échantillon de 1846 entreprises industrielles françaises prises sur la période 1999-2006, à l'aide d'une étude sur panel dynamique en utilisant la méthode des moments généralisée (GMM), que la structure du capital n'a aucune influence sur la profitabilité des entreprises françaises quelle que soit la taille de l'entreprise.

Mots-clés : Structure du capital, profitabilité, GMM, données de panel.

# Capital structure and profitability
# The case of French industrial firms

Abstract

The objective of this article is to analyze the impact of capital structure on profitability. This impact can be explained by three essential theories: signaling theory, tax theory and the agency costs theory. A sample of 1846 French industrial firms are taken over the period 1999-2006, as a dynamic panel study by using the generalized method of moments (GMM). We show that capital structure has no influence on the profitability of French firms, regardless the size of the company.

---

[*] Laboratoire d'Économie d'Orléans (LEO), Université d'Orléans. Faculté de Droit, d'Économie et de Gestion. Rue de Blois - BP : 6739, 45067 Orléans Cedex 2. E-mail : mazen.kebewar@etu.univ-orleans.fr.

∗ Department of Statistics and Management Information Systems, University of Aleppo, Faculty of Economics. E-mail: mazen.kebewar@gmail.com {مازن كبه وار، جامعة حلب، كلية الاقتصاد – قسم الاحصاء ونظم المعلومات الادارية}.

# 1. Introduction

Le rôle de la structure du capital dans l'explication de la performance des entreprises fait l'objet de plusieurs recherches depuis plus de cinquante ans (Modigliani et Miller 1958). Cependant ce rôle reste un sujet d'actualité qui attire l'attention de beaucoup de chercheurs comme, Goddard et al. (2005), Berger et Bonaccorsi (2006), Rao et al. (2007), Baum et al. (2007), Weill (2008), Nunes et al. (2009), Margaritis et Psillaki (2010) et Kebewar (2012).

En effet, les chercheurs analysent la structure du capital et essayent de déterminer si une structure du capital optimale existe. La structure du capital optimale est généralement définie comme celle qui minimise les coûts de capital d'entreprise, tout en maximisant la valeur de l'entreprise. Autrement dit, la structure du capital optimale est celle qui maximise la profitabilité d'entreprise.

Par ailleurs, le désaccord entre chercheurs s'observe sur le plan théorique. Il existe trois théories essentielles qui peuvent mettre en évidence l'influence de l'endettement sur la profitabilité des entreprises, à savoir : la théorie du signal, la théorie de l'agence et l'influence de la fiscalité. D'abord, selon la théorie de signal, l'endettement, en situation d'information asymétrique, devrait être positivement corrélé avec la profitabilité. D'après la théorie de l'agence, il existe deux effets contradictoires de l'endettement sur la profitabilité, le premier effet est positif dans le cas des coûts de l'agence des fonds propres entre actionnaires et dirigeants, mais, le deuxième effet est négatif, il résulte des coûts d'agence des dettes financières entre actionnaires et prêteurs. Enfin, l'influence de la fiscalité est plutôt complexe et difficile à prédire car elle dépend du principe de déductibilité fiscale des intérêts des dettes, de l'imposition sur le revenu et des déductions d'impôt non liées à l'endettement.

De plus, le désaccord entre chercheurs s'observe non seulement sur le plan théorique, mais aussi sur le plan empirique. Un effet négatif de l'endettement sur la profitabilité a été confirmé par Majumdar et Chhibber (1999), Eriotis et al. (2002), Ngobo et Capiez (2004), Goddard et al. (2005), Rao et al. (2007), Zeitun et Tian (2007) et Nunes et al. (2009). Par contre, Baum et al. (2006), Berger et Bonaccorsi (2006), Margaritis et Psillaki (2007), Baum et al. (2007) et Margaritis et Psillaki (2010), ils ont montré une influence positive. De plus, Simerly et LI (2000), Mesquita et Lara (2003) et Weill (2008), ils ont trouvé les deux effets dans leurs études. Par ailleurs, Berger et Bonaccorsi (2006), Margaritis et Psillaki (2007) et Kebewar (2012) ont trouvé la présence d'un effet non linéaire (une courbe de forme inverse de U). Enfin, un effet non significatif a été confirmé par Baum et al (2007) sur des entreprises industrielles américaines.

La contradiction des résultats empiriques peut être expliquée par plusieurs facteurs. D'abord, ces études empiriques portent sur des types d'échantillons différents (pays, secteurs, entreprises et périodes). De plus, les chercheurs ont utilisé différents mesures de la profitabilité comme une variable dépendante[1] et différents ratios d'endettement en tant que variable indépendante[2]. Enfin, ces travaux ont appliqué différentes méthodologies[3].

L'objectif de cet article est d'analyser l'effet de la structure du capital sur la profitabilité des entreprises françaises. L'importance de ce sujet est que l'endettement est un choix risqué dont les conséquences sur la performance de l'entreprise peuvent être considérables (par exemple le risque de faillite et ses conséquences pour les parties prenantes).

En effet, la littérature empirique concernant l'incidence de la structure du capital sur la profitabilité nous conduit à faire deux constats. Le premier constat est que la plupart des études empiriques, en la matière, se sont concentré sur les entreprises cotées. Le deuxième constat est lié à la rareté des études sur les entreprises françaises, car il y en a quelque études comme : Goddard et al. (2005), Weill (2008), Margaritis et Psillaki (2010) et récemment Kebewar (2012). Ces deux constats ont motivé notre étude. Donc, nous allons essayer de trouver empiriquement l'effet de l'endettement sur la profitabilité des entreprises françaises de type anonymes et de type SARL qui ne sont pas cotées. De plus, et dans le but d'améliorer la précision de l'estimation en réduisant l'hétérogénéité entre les différents tailles, nous allons étudier le comportement de ces entreprises suivant leur taille. D'ailleurs, nous allons analyser non seulement l'effet linéaire de la structure du capital sur la profitabilité, mais aussi l'effet non linéaire en estimant un modèle quadratique qui prend en compte la variable d'endettement au carré dans l'équation de la régression.

Pour ce faire, nous allons mettre en place la méthode des moments généralisée (GMM) sur un échantillon de (1846) entreprises industrielles observés sur la période (1999-2006), ces entreprises sont réparties sur trois tailles d'entreprise (TPE, PME et ETI). Selon les défenseurs de la méthode de (GMM), elle permet d'apporter des solutions aux problèmes de biais de simultanéité, de causalité inverse (surtout entre l'endettement et la profitabilité) et des éventuelles variables omises.

---

[1] ROA, ROE, ROI, PROF, *Tobin's Q*, Bénéfice sur ventes, Performance commerciale, VRS : *Technical Efficiency*, CRS : *Technical Efficiency*, *Profit Margin*, *Frontier efficiency*, BTI : le ratio résultat avant intérêts et impôts sur l'actif total et d'autres.
[2] Ratio de dette totale, ratio de dette à CT, ratio de dette à LT et d'autres.
[3] MCO, MCG, doubles moindres carrés, Moindres carrés pondérés, Effet fixe, Effet aléatoire, le modèle de la décomposition de la variance, le modèle de covariance, Maximum de vraisemblance, Méthode des équations simultanées, Régression quantile et GMM (qui est le plus récente et le plus utilisée).

La suite de cet article est structurée de la façon suivante : d'abord, nous abordons, les caractéristiques de l'échantillon et les variables. Puis, nous montrons la méthodologie. Ensuite, nous présentons les résultats empiriques. Enfin, nous terminons cet article par une conclusion générale.

## 2. Les données

### 2.1 La description des données

L'échantillon, qui est tiré de la base de données (Diane), est constitué d'un panel non cylindré de 1846 entreprises industrielles françaises sur la période de 1999 à 2006, ces entreprises sont des sociétés de type anonymes et de SARL non cotées appartenant à trois classes de taille[4] (TPE, PME et ETI).

A cause de la particularité de leur politique d'endettement, les entreprises publiques seront exclues de l'étude. Seront également exclues les entreprises dont les capitaux propres sont négatifs. De plus, les observations aberrantes ont été supprimées suivant la procédure de Kremp (1995)[5]. Comme l'estimation se fait à l'aide de la méthode des moments généralisée (GMM) qui nécessite beaucoup de données, les grandes entreprises, peu nombreuses, n'ont pas été prises en compte.

Alors, le tableau (1) montre les statistiques descriptives et la répartition des entreprises selon les tailles[6]. Le calcul des variables utilisées dans notre analyse est fourni en annexe.

**Tableau (1)**
STATISTIQUES DESCRIPTIVES DES PRINCIPALES VARIABLES

| CLASSE DE TAILLE | CLASSE (1) 1 -19 | | CLASSE (2) 20 - 249 | | CLASSE (3) 250 - 4999 | | TOTAL | |
|---|---|---|---|---|---|---|---|---|
| Nombre d'entreprises | 952 | | 747 | | 147 | | 1846 | |
| Nombre d'observations | 6337 | | 5136 | | 909 | | 12382 | |
| Prof1 | 0,085 | (0,084) | 0,063 | (0,066) | 0,065 | (0,058) | 0,074 | (0,076) |
| Prof2 | 0,091 | (0,082) | 0,069 | (0,062) | 0,065 | (0,058) | 0,080 | (0,074) |
| ROA | 0,064 | (0,067) | 0,046 | (0,052) | 0,050 | (0,045) | 0,055 | (0,060) |
| Dt | 0,582 | (0,180) | 0,569 | (0,168) | 0,556 | (0,159) | 0,575 | (0,173) |
| Gar2 | 0,160 | (0,131) | 0,168 | (0,124) | 0,179 | (0,115) | 0,165 | (0,127) |
| Impot1 | 0,187 | (0,128) | 0,209 | (0,162) | 0,253 | (0,689) | 0,201 | (0,233) |
| Crois | 0,058 | (0,198) | 0,044 | (0,167) | 0,051 | (0,214) | 0,052 | (0,187) |

\* Les valeurs entre parenthèses sont les écart-type.

---

[4] Selon la classification de l'INSEE, les toutes petites entreprises (TPE) dont l'effectif est inférieur à 20 employés, les petites et moyennes entreprises (PME) ayant un effectif entre 20 et 249 employés et enfin, les entreprises de taille intermédiaire (ETI) dont l'effectif est compris entre 250 et 4999 employés.
[5] Nous avons supprimé les observations qui se situent hors de l'intervalle défini par les premier et troisième quartiles plus ou moins cinq fois l'écart interquartile.
[6] La composition de l'échantillon est disponible à la demande de l'auteur.

## 2.2 Mesures des variables

### 2.2.1 Variable dépendante (la profitabilité)

En théorie, la profitabilité d'entreprise peut se mesurer de différentes manières. Dans notre étude et pour comparer nos résultats, nous allons utiliser trois mesures de la profitabilité : Prof1, Prof2 et ROA. (Prof1) est mesurée en divisant le résultat d'exploitation par l'actif total. (Prof2) se calcule en rapportant le résultat avant intérêts et impôt à l'actif total. La rentabilité économique (ROA) est mesurée en rapportant le résultat d'exploitation (auquel les impôts sur les bénéfices sont soustraits) à l'ensemble des capitaux.

### 2.2.2 Variables explicatives

- *L'endettement*

Selon la littérature le ratio d'endettement peux se mesurer par plusieurs méthodes. Le ratio d'endettement total, le ratio d'endettement à court, moyen et long terme. Dans la cadre de notre étude, nous avons défini le ratio d'endettement total en divisant la somme des dettes à court et à long terme par l'actif total.

- *La garantie*

La garantie a deux effets contradictoires sur la profitabilité. D'un coté, nous attendons un effet positif selon Himmelberg et al. (1999). Ils montrent que les immobilisations corporelles sont faciles à surveiller et fournir des bonnes garanties, donc, ils ont tendance à atténuer les conflits d'agence de la dette entre actionnaires et créditeurs. D'un autre coté, nous devons prévoir, Selon Deloof (2003) et Nucci et al. (2005), une corrélation négative car les entreprises ayant des niveaux élevés d'immobilisations corporelles ont tendance à être moins rentables, car les entreprises avec des niveaux élevés d'actifs incorporelles (sous forme de liquidité) ont plus les possibilités d'investissement à long terme, à l'innovation et aux recherches et développement (R&D).

La relation négative entre la garantie et la profitabilité a été confirmé dans un certain nombre d'études comme Rao et al (2007), Zeitun et Tian (2007), Weill (2008) et Nunes et al (2009). Par contre, les travaux de Majumdar et Chhibber (1999) et Margaritis et Psillaki (2007) trouvent des relations positives.

Pour savoir l'effet de la garantie sur la profitabilité des entreprises, nous avons retenu le ratio (Gar), il se calcule en rapportant la somme des immobilisations corporelles nettes à l'actif total.

- *L'impôt*

L'impact de l'impôt sur la profitabilité d'une entreprise est plutôt complexe et difficile à prévoir. Parce que cela dépond du principe de déductibilité fiscale des intérêts des dettes. C'est-à-dire que si une entreprise ne profite pas de ce principe, nous prévoyons un effet négatif de l'impôt sur la profitabilité. Au contraire, si une entreprise en profite, cet impact devrait être positif ou pas significatif. Zeitun et Tian (2007) ont mis en évidence un effet positif de l'impôt sur la profitabilité.

L'influence de la fiscalité sur la profitabilité des entreprises sera mettre en évidence par l'utilisation du ratio de l'impôt dans l'équation de régression. Ce ratio (Impot) a été calculé en divisant l'impôt payé sur le bénéfice avant intérêt et impôt.

- *Les opportunités de croissance*

Il est prévu que les entreprises ayant des opportunités de croissance élevées ont un taux de rendement élevé, car ces entreprises sont capables de générer des bénéfices de l'investissement. Donc, les opportunités de croissance d'une entreprise devraient influencer positivement sur sa performance.

L'influence positive des opportunités de croissance sur la profitabilité a été confirmée par la plupart des travaux empiriques tels que Margaritis et Psillaki (2007), Zeitun et Tian (2007) et Nunes et al. (2009). Par contre, Margaritis et Psillaki (2010) ont trouvé un effet négative seulement dans le secteur de chimie en France.

Plusieurs mesures de l'opportunité de croissance des entreprises existent dans la littérature. Mais, dans le cadre de notre analyse, nous allons utiliser le ratio de l'opportunité de croissance (Crois) qui est mesurée par la variation du total de l'actif d'une année sur l'autre. Ce ratio doit nous permettre d'évaluer l'influence de la dynamique de croissance de l'entreprise sur sa profitabilité.

## 3. La méthodologie

Le modèle à estimer pour analyser l'impact de l'endettement sur la profitabilité des entreprises se présente sous la forme suivante :

$$PROF_{i,t} = \beta_0 + \beta_1 D_{i,t} + \beta_2 GAR_{i,t} + \beta_3 IMPOT_{i,t} + \beta_4 CROIS_{i,t} + \sum_{n=1}^{9} \beta_n dumt_n + \eta_i + \varepsilon_{it}$$

Où (i) désigne l'entreprise étudiée et (t) fait référence à la période d'analyse. La variable dépendante du modèle est le ratio de profitabilité (Prof1, Prof2 ou ROA). Par ailleurs (D), (GAR), (IMPOT), et (CROIS) représentent respectivement les ratios de l'endettement, de la garantie, de l'impôt et de l'opportunité de croissance.

L'influence du temps est prise en compte par l'introduction d'indicatrices temporelles annuelles (dumt) qui captent l'effet spécifique des années (1999-2007). L'effet individuel fixe relatif aux entreprises est représenté par le terme ($\eta_i$). Enfin le terme d'erreur, qui est supposé indépendant et identiquement distribué (*i.i.d*), est représenté par le terme ($\varepsilon_{it}$).

En ce qui concerne l'analyse de la non-linéarité de l'impact de la dette sur la profitabilité, nous allons estimer un modèle quadratique qui prend en compte la variable d'endettement au carré dans l'équation de la régression. Donc, le modèle à estimer dans ce cadre devient sous la forme suivante :

$$PROF_{i,t} = \beta_0 + \beta_1 D_{i,t} + \beta_2 D\text{\^}2_{i,t} + \beta_3 GAR_{i,t} + \beta_4 IMPOT_{i,t} + \beta_5 CROIS_{i,t} + \sum_{n=1}^{9} \beta_n dumt_n + \eta_i + \varepsilon_{it}$$

L'hypothèse nulle du test de linéarité consiste à tester : (H0 : $\beta_2 = 0$). Si cette hypothèse est rejetée, nous pouvons conclure à l'existence de la non-linéarité entre l'endettement et la profitabilité. Selon la théorie de l'agence, l'effet de l'endettement sur la profitabilité doit être positif lorsque ($\beta_1 > 0$ et $\beta_1 + 2\beta_2 D_{i,t} > 0$). Toutefois, si le ratio d'endettement arrive à un niveau suffisamment élevé, cet effet peut être négatif. Alors, la spécification quadratique est compatible avec la possibilité que la relation entre l'endettement et la profitabilité ne peut pas être monotone, elle peut passer du positif au négatif à un niveau d'endettement élevé. Le ratio de la dette aura un impact négatif sur la profitabilité lorsque ($D_{i,t} < -\beta_1 / 2\beta_2$). Autrement dit, la condition suffisante, pour que la relation entre l'endettement et la profitabilité soit en forme de (*cloche*), est que $\beta_2 < 0$.

Nous pouvons soupçonner des problèmes d'endogéneité au niveau de l'équation d'estimation liés à une causalité des variables exogènes vers la variable dépendante (plus particulièrement la variable de l'endettement). Donc, les méthodes économétriques traditionnelles comme (MCO, effet fixe et moindres carrés quasi-généralisés) ne nous permettent pas d'obtenir des estimations efficientes d'un tel modèle. Alors, pour résoudre ce problème, nous allons mettre en place la méthode des moments généralisée sur panel (GMM) proposée par Arellano et Bond (1991), et développée plus tard par Arellano et Bover (1995) et Blundell et Bond (1998). Selon les défenseurs de cette méthode, elle permet d'apporter des solutions aux problèmes de biais de simultanéité, de causalité inverse (surtout entre l'endettement et la profitabilité) et des éventuelles variables omises. D'ailleurs, elle contrôle les effets spécifiques individuels et temporels.

En effet, la méthode des (GMM) permet de régler le problème d'endogénéité non seulement au niveau de la variable de l'endettement, mais aussi au niveau des autres variables explicatives par l'utilisation d'une série de variables instrumentales générées par les retards des variables.

De plus, il faut ajouter que la méthode des (GMM) sur panel présente un autre avantage, elle génère les instruments à partir des variables explicatives; ce qui n'est pas le cas des autres méthodes traditionnelles de variables instrumentales comme (2SLS et 3SLS), qui nécessitent le choix de variable instrumentale théorique corrélée avec les variables explicatives et non corrélée avec le résidu, ce qui est difficile à trouver.

La mise en place de la méthode des GMM s'effectue en utilisant la procédure (XTABOND2)[7] sur le logiciel (STATA). Le modèle sera estimé par la méthode des moments généralisée en système et en deux étapes. Dans le but de choisir la meilleure spécification de modèle, nous avons examiné plusieurs spécifications suivant différentes hypothèses concernant l'endogénéité des variables.

## 4. Les résultats

### 4.1 L'analyse descriptive

Nous remarquons selon le tableau (2) qu'il y a une diminution de la profitabilité des entreprises françaises. Le ratio de profitabilité a été diminué entre 3 et 11 points (selon le ratio utilisé Prof1, Prof2 ou ROA).

Le ratio de la profitabilité est passé de (0,088) en 1999 à (0,077) en 2006 en utilisant la mesure Prof1, de (0,93) à (0,083) pour la mesure Prof2 et de (0,061) à (0,059) pour ROA.

---

[7] Pour plus de détails, Roodman, D. (2006), page 54.

**Tableau (2)**
L'EVOLUTION DU RATIO DE LA PROFITABILITE

| ANNEE | Prof1 | Prof2 | ROA | (Prof1) | | | (Prof2) | | | (ROA) | | |
|---|---|---|---|---|---|---|---|---|---|---|---|---|
| | | | | TPE | PME | ETI | TPE | PME | ETI | TPE | PME | ETI |
| 1999 | 0,088 | 0,093 | 0,061 | 0,102 | 0,076 | 0,069 | 0,108 | 0,080 | 0,067 | 0,070 | 0,054 | 0,053 |
| 2000 | 0,090 | 0,094 | 0,065 | 0,105 | 0,076 | 0,059 | 0,110 | 0,081 | 0,063 | 0,075 | 0,056 | 0,044 |
| 2001 | 0,080 | 0,084 | 0,060 | 0,092 | 0,068 | 0,062 | 0,096 | 0,072 | 0,063 | 0,071 | 0,050 | 0,048 |
| 2002 | 0,067 | 0,074 | 0,051 | 0,078 | 0,056 | 0,060 | 0,085 | 0,063 | 0,058 | 0,060 | 0,040 | 0,046 |
| 2003 | 0,070 | 0,075 | 0,053 | 0,079 | 0,060 | 0,068 | 0,085 | 0,065 | 0,066 | 0,061 | 0,044 | 0,054 |
| 2004 | 0,064 | 0,070 | 0,049 | 0,070 | 0,059 | 0,060 | 0,077 | 0,063 | 0,060 | 0,054 | 0,042 | 0,047 |
| 2005 | 0,067 | 0,072 | 0,052 | 0,076 | 0,056 | 0,067 | 0,083 | 0,061 | 0,067 | 0,059 | 0,042 | 0,054 |
| 2006 | 0,077 | 0,083 | 0,059 | 0,086 | 0,064 | 0,081 | 0,095 | 0,069 | 0,080 | 0,067 | 0,048 | 0,065 |
| changement | -0,011 | -0,009 | -0,003 | -0,015 | -0,012 | 0,012 | -0,013 | -0,011 | 0,012 | -0,003 | -0,006 | 0,012 |

Note : (TPE) : moins de 19 personnes, (PME) entre 20 et 249 personnes et (ETI) : entre 250 et 4999 personnes.

En ce qui concerne l'évolution de la profitabilité selon les tailles, nous constatons que les très petites entreprises (TPE) réalisent une profitabilité plus importante que des entreprises de tailles (PME) et (ETI), c'est-à-dire, il existe une relation décroissante et stable sur la période entre la taille et la profitabilité.

### 4.2 La corrélation entre les variables

Le tableau ci-dessous présente les coefficients de corrélation entre les différentes variables utilisées dans notre modèle. Prenant en compte les rapports entre la variable d'endettement (Dt) et les variables dépendantes (Prof1, Prof2 et ROA), nous constatons que l'endettement révèle être négativement corrélée avec la profitabilité, mais, il semple que l'effet négatif est très faible. D'ailleurs, la variable de garantie (Gar) est aussi corrélée négativement avec la profitabilité. Par contre, en ce qui concerne la variable de l'impôt (Impot) et celle de l'opportunité de croissance (Crois), il parait qu'elles ont une corrélation positive avec la profitabilité. En regardant les relations entre les variables indépendantes elles-mêmes, les résultats[8] révèlent que la multicollinériaté n'est pas un problème pour l'application des techniques d'analyse.

**Tableau (3)**
LA CORRELATION DE PEARSON ENTRE LES VARIABLES DE REGRESSION

| | Prof1 | Prof2 | ROA | Dt | Gar | Impot | VIF |
|---|---|---|---|---|---|---|---|
| Prof2 | 0.929*** | 1 | | | | | |
| ROA | 0.975*** | 0.868*** | 1 | | | | |
| Dt | -0.052*** | -0.087*** | -0.0065 | 1 | | | 1,02 |
| Gar | -0.094*** | -0.084*** | -0.092*** | 0.081*** | 1 | | 1,01 |
| Impot | 0.240*** | 0.234*** | 0.181*** | -0.113*** | -0.053*** | 1 | 1,03 |
| Crois | 0.198*** | 0.173*** | 0.203*** | 0.055*** | -0.0075 | 0.054*** | 1,01 |

Note : *** significatif au seuil d'erreur de 1%.

---

[8] Selon le test de VIF (Variance Inflation Facteurs).

## 4.3 L'analyse économétrique

Nous avons estimé l'effet de la structure du capital sur la profitabilité des entreprises industrielles françaises, en utilisant différents représentants de la profitabilité de l'entreprise comme (Prof1, Prof2 et ROA). D'ailleurs, nous avons utilisé deux modèles différents (linéaire et non linéaire) pour vérifier la présence d'un non linéarité de cet impact. De plus, l'estimation a été détaillée en étudiant spécifiquement le comportement des entreprises selon leur taille (TPE, PME et ETI). Donc, le tableau (4) présente tous les résultats obtenus.

Concernant l'efficacité de l'estimateur de GMM en panel, nous pouvons dire que tous nos résultats sont robustes pour les raisons suivantes. D'abord, les instruments utilisés dans nos régressions sont valides, car le test de Hansen ne permet pas de rejeter l'hypothèse de validité des variables retardés en niveau et en différence comme instruments. De plus, nous remarquons qu'il n'y a pas d'autocorrélation de second ordre des erreurs de l'équation en différence (AR2), car, le test d'autocorrélation de second ordre d'Arellano et Bond ne permette pas de rejeter l'hypothèse d'absence d'autocorrélation de second ordre[9].

A propos de l'impact de la structure du capital sur la profitabilité, nous remarquons que la structure du capital n'a aucune influence sur la profitabilité, ni de façon linéaire, ni de façon non linéaire. De plus, lorsque nous détaillons l'analyse en utilisant différents classes de taille, nous constatons aussi qu'il n'y a pas d'impact de la structure du capital sur la profitabilité quelle que soit la classe de taille de l'entreprise.

En ce qui concerne les variables de contrôle, nous remarquons, d'abord, que la variable de la garantie (Gar) affecte négativement la profitabilité, cela signifie que les entreprises investissent trop dans les immobilisations d'une manière qui n'améliore pas leurs performances, ou elles n'utilisent pas leurs immobilisations efficacement. Par contre, Il est remarqué que l'opportunité de croissance (Crois) et l'impôt (Impot) affectent positivement la profitabilité, cela signifie que les entreprises réalisent plus de profitabilité lorsqu'elles ont plus des opportunités de croissance et d'impôts.

---

[9] Dans un souci de robustesse, nous avons estimé le modèle par la méthode de GMM en deux étapes et une étape, et par conséquent, nous avons obtenue les mêmes résultats.

**Tableau (4)**
L'IMPACT DE LA STRUCTURE DE CAPITAL SUR LA PROFITABILITE

**VARIABLE DEPENDANTE (PROF1)**

|           | Ensemble (Prof1) |           | CLASSE (1) |           | CLASSE (2) |           | CLASSE (3) |           |
|-----------|------------------|-----------|------------|-----------|------------|-----------|------------|-----------|
| Dt        | -0,393           | 0,682     | 0,348      | -0,536    | 0,069      | -1,482    | -0,056     | 0,096     |
|           | (-1,21)          | (0,42)    | (1,37)     | (-0,48)   | (0,64)     | (-1,22)   | (-0,67)    | (0,14)    |
| Dt*2      |                  | -0,648    |            | 0,507     |            | 1,482     |            | -0,120    |
|           |                  | (-0,45)   |            | (0,49)    |            | (1,35)    |            | (-0,21)   |
| Gar       | -0,069***        | -0,067*** | -0,050**   | -0,051*   | -0,030**   | -0,010    | -0,078***  | -0,073*** |
|           | (-3,95)          | (-2,87)   | (-2,22)    | (-1,95)   | (-2,19)    | (-0,41)   | (-2,85)    | (-2,62)   |
| Impot     | 0,003            | 0,054**   | 0,481***   | 0,300***  | 0,149***   | 0,191***  | 0,004      | 0,004     |
|           | (0,05)           | (2,14)    | (3,05)     | (3,10)    | (4,98)     | (5,24)    | (0,95)     | (1,04)    |
| Crois     | 0,071***         | 0,069***  | 0,068***   | 0,093***  | 0,062***   | 0,060***  | 0,005      | 0,005     |
|           | (3,93)           | (3,50)    | (2,73)     | (5,63)    | (6,06)     | (4,44)    | (0,83)     | (0,74)    |
| Constant  | 0,263*           | -0,088    | -0,154     | 0,167     | 0,005      | 0,373     | 0,120**    | 0,075     |
|           | (1,70)           | (-0,21)   | (-1,06)    | (0,60)    | (0,09)     | (1,21)    | (2,14)     | (0,41)    |
| Observations | 10536         | 10536     | 5385       | 5385      | 4389       | 4389      | 762        | 762       |
| Nombre de firme | 1846       | 1846      | 952        | 952       | 747        | 747       | 147        | 147       |
| p-value sargan | 0,73        | 0,55      | 0,91       | 0,49      | 0,08       | 0,79      | 0,86       | 0,91      |
| P-value AR(2) | 0,10         | 0,13      | 0,98       | 0,59      | 0,10       | 0,11      | 0,78       | 0,81      |

**VARIABLE DEPENDANTE (PROF2)**

|           | Ensemble (Prof2) |           | CLASSE (1) |           | CLASSE (2) |           | CLASSE (3) |           |
|-----------|------------------|-----------|------------|-----------|------------|-----------|------------|-----------|
| Dt        | -0,291           | 0,330     | 0,316      | -1,325    | 0,126      | -1,299    | 0,016      | -0,437    |
|           | (-1,01)          | (0,21)    | (1,29)     | (-0,92)   | (1,21)     | (-1,14)   | (0,21)     | (-0,63)   |
| Dt*2      |                  | -0,318    |            | 1,228     |            | 1,366     |            | 0,399     |
|           |                  | (-0,22)   |            | (0,95)    |            | (1,31)    |            | (0,64)    |
| Gar       | -0,059***        | -0,055**  | -0,037*    | -0,025    | -0,029**   | -0,009    | -0,079***  | -0,072**  |
|           | (-4,13)          | (-2,44)   | (-1,72)    | (-0,85)   | (-2,06)    | (-0,38)   | (-2,88)    | (-2,46)   |
| Impot     | 0,016            | 0,057**   | 0,474***   | 0,342***  | 0,156***   | 0,196***  | 0,001      | 0,000     |
|           | (0,34)           | (2,03)    | (3,10)     | (3,47)    | (5,45)     | (5,23)    | (0,26)     | (0,13)    |
| Crois     | 0,060***         | 0,058***  | 0,057**    | 0,074***  | 0,050***   | 0,049***  | 0,003      | 0,005     |
|           | (3,83)           | (3,36)    | (2,38)     | (4,36)    | (5,37)     | (3,87)    | (0,54)     | (0,75)    |
| Constant  | 0,221            | 0,005     | -0,130     | 0,364     | -0,018     | 0,318     | 0,088*     | 0,201     |
|           | (1,61)           | (0,01)    | (-0,93)    | (1,01)    | (-0,33)    | (1,10)    | (1,68)     | (1,13)    |
| Observations | 10536         | 10536     | 5385       | 5385      | 4389       | 4389      | 762        | 762       |
| Nombre de firme | 1846       | 1846      | 952        | 952       | 747        | 747       | 147        | 147       |
| p-value sargan | 0,69        | 0.60      | 0,82       | 0,71      | 0,05       | 0,48      | 0,96       | 0,74      |
| P-value AR(2) | 0,12         | 0,13      | 0,80       | 0,77      | 0,09       | 0,16      | 0,18       | 0,13      |

**VARIABLE DEPENDANTE (ROA)**

|           | Ensemble (ROA)   |           | CLASSE (1) |           | CLASSE (2) |           | CLASSE (3) |           |
|-----------|------------------|-----------|------------|-----------|------------|-----------|------------|-----------|
| Dt        | -0,307           | 0,505     | 0,292      | -0,202    | 0,081      | -1,098    | -0,056     | 0,283     |
|           | (-1,14)          | (0,38)    | (1,36)     | (-0,21)   | (0,88)     | (-1,13)   | (-0,83)    | (0,54)    |
| Dt*2      |                  | -0,494    |            | 0,171     |            | 1,129     |            | -0,282    |
|           |                  | (-0,42)   |            | (0,19)    |            | (1,28)    |            | (-0,61)   |
| Gar       | -0,056***        | -0,056*** | -0,041**   | -0,047**  | -0,030***  | -0,015    | -0,063***  | -0,059**  |
|           | (-3,94)          | (-2,92)   | (-2,18)    | (-2,14)   | (-2,63)    | (-0,75)   | (-2,80)    | (-2,27)   |
| Impot     | -0,008           | 0,030*    | 0,354***   | 0,175*    | 0,097***   | 0,131***  | 0,001      | 0,002     |
|           | (-0,20)          | (1,72)    | (2,67)     | (1,92)    | (3,85)     | (4,54)    | (0,58)     | (0,80)    |
| Crois     | 0,059***         | 0,057***  | 0,057***   | 0,084***  | 0,051***   | 0,050***  | 0,006      | 0,005     |
|           | (3,97)           | (3,64)    | (2,73)     | (5,37)    | (6,05)     | (4,56)    | (1,26)     | (0,96)    |
| Constant  | 0,205            | -0,059    | -0,127     | 0,092     | -0,010     | 0,270     | 0,097**    | 0,004     |
|           | (1,59)           | (-0,17)   | (-1,04)    | (0,40)    | (-0,20)    | (1,09)    | (2,05)     | (0,03)    |
| Observations | 10536         | 10536     | 5385       | 5385      | 4389       | 4389      | 762        | 762       |
| Nombre de firme | 1846       | 1846      | 952        | 952       | 747        | 747       | 147        | 147       |
| p-value sargan | 0,65        | 0,53      | 0,94       | 0,399     | 0,12       | 0,781     | 0,72       | 0,842     |
| P-value AR(2) | 0,25         | 0,21      | 0,80       | 0,50      | 0,15       | 0,11      | 0,82       | 0,856     |

Note : Le modèle utilisé : GMM en système, deux étapes, instruments : la variable d'endettement (Dt) retardée t-3, la variable d'endettement au carré (Dt*2) retardée t-6, le reste des variables explicatives sont exogènes.

# 5. Conclusion

Dans cet article, nous nous sommes intéressés à l'effet de la structure du capital sur la profitabilité des entreprises françaises. Autrement dit, cet article avait pour objectif d'élargir le champ de la connaissance sur l'influence de l'endettement sur la profitabilité.

En effet, il y a trois théories essentielles qui peuvent mettre en évidence l'influence de l'endettement sur la profitabilité des entreprises à savoir : la théorie du signal, la théorie de l'agence et l'influence de la fiscalité. De plus, le désaccord entre chercheurs s'observe non seulement sur le plan théorique, mais aussi sur le plan empirique.

La rareté des études sur les entreprises françaises et la concentration des études sur les entreprises cotées ont motivé notre étude. Pour cela, nous avons examiné empiriquement cet impact en utilisant la méthode d'estimation de GMM sur un panel non cylindré de 1846 entreprises industrielles françaises de type anonymes et de type SARL, qui ne sont pas cotées, sur une période de huit ans (1999-2006). Ces entreprises sont réparties sur trois tailles d'entreprise différentes (TPE, PME et ETI), cela a permis d'améliorer la précision de l'estimation en réduisant l'hétérogénéité entre les différentes tailles d'entreprises. Par ailleurs, nous avons analysé non seulement l'effet linéaire de la structure du capital sur la profitabilité, mais aussi l'effet non linéaire en estimant un modèle quadratique qui prend en compte la variable d'endettement au carré dans l'équation de la régression.

À l'issue de cette étude, nous pouvons souligner que la structure du capital n'a aucune influence sur la profitabilité, ni de façon linéaire, ni de façon non linéaire. De plus, lorsque nous détaillons l'analyse en utilisant différents classes de taille, nous constatons aussi qu'il n'y a pas d'impact quelle que soit la classe de taille de l'entreprise.

Dans des recherches futures, il serait intéressant de prendre en compte quelques réflexions. D'abord, il sera d'intérêt à étendre cette analyse à travers différentes composantes de l'endettement des entreprises (la dette à long terme et celle à court terme), c'est parce que, selon la plupart des études, des effets contradictoires ont été trouvé. Ensuite, il faudrait ajouter des nouvelles variables spécifiques à l'entreprise et au secteur, entre autre, la structure de propriété du capital d'entreprise et l'environnement dans lequel les entreprises évoluent. Enfin, compte tenu du fait que la relation entre la structure du capital et la profitabilité peux être non linéaire, nous pourrons toutefois approfondir le présent article en utilisant des méthodes économétriques qui peuvent analyser l'effet non linéaire comme la Régression quantile et le modèle à changement de régime.

# Annexe

$Dt_{i,t}$ (Ratio d'endettement) : Le ratio de total de dettes (à court et à long terme) par rapport au total de l'actif.

$$Dt_{i,t} = \frac{DS + DT + DU + DV + DW + DX + DY + DZ + EA}{EE}$$

    DS : Emprunts obligataires convertibles.
    DT : Autres emprunts obligataires.
    DU : Emprunts et dettes auprès des établissements de crédit.
    DV : Emprunts et dettes financières divers.
    DW : Avances et acomptes reçus sur commandes en cours.
    DX : Dettes fournisseurs et comptes rattachés.
    DY : Dettes fiscales et sociales.
    DZ : Dettes sur immobilisations et comptes rattachés.
    EA : Autres dettes.
    EE : Total du passif (total général de bilan).

$CROIS_{i,t}$ (Croissance) : La variation du total de l'actif d'une année sur l'autre.

$$CROIS_{i,t} = \frac{EE_t - EE_{t-1}}{EE_{t-1}}$$

$IMPOT_{i,t}$ (Impôt) : Le ratio de l'impôt payé sur le bénéfice avant intérêt et impôt.

$$IMPOT_{i,t} = \frac{HK}{HN + HK + GR}$$

    HK : Impôts sur les bénéfices.
    HN : Bénéfice ou perte.
    GR : Intérêts et charges assimilées.

$GAR_{i,t}$ (Garantie): La somme des immobilisations corporelles nets divisé par le l'actif total.

$$GAR_{i,t} = \frac{(ANN + APN + ARN + ATN + AVN + AXN)}{EE}$$

    ANN : Terrains nets.
    APN : Constructions nettes.
    ARN : Installations techniques, matériel et outillage industriels nets.
    ATN : Autres immobilisations corporelles nettes.
    AVN : Immobilisations en cours nettes.
    AXN : Avance et acomptes nettes.

$PROF1_{i,t}$ (Profitabilité1): Le résultat d'exploitation sur l'actif total. $\quad PROF1_{i,t} = \frac{GG}{EE}$

    GG : Résultat d'exploitation.

$PROF2_{i,t}$ (Profitabilité2): Le résultat avant intérêts et impôts sur l'actif total.

$$PROF2_{i,t} = \frac{HN + HK + GR}{EE}$$

    HN : Bénéfice ou perte.
    HK : Impôts sur les bénéfices
    GR : Intérêts et charges assimilées.

$ROA_{i,t}$ *(Return On Assets):* La rentabilité économique est égale à : $\quad ROA_{i,t} = \frac{GG - HK}{CP + DT}$

    GG : Résultat d'exploitation.
    HK : Impôts sur les bénéfices.